%% file: main.tex
\documentclass[10pt,sigconf,letterpaper]{acmart}

\usepackage{tikz}
\usepackage{amsmath}
\usepackage{booktabs}
\usepackage{graphicx}
\usepackage{tablefootnote}
\usepackage{multirow}
\usepackage{url}
\usepackage{enumitem}
\usepackage{pifont}
\usepackage{framed}
\usepackage{colortbl}
\definecolor{mygray}{gray}{.9}
\definecolor{codegreen}{rgb}{0,0.6,0}
\definecolor{codepurple}{rgb}{0.58,0,0.82}
\usepackage{listings}
\graphicspath{{images/}}
 
\newcommand{\code}[1]{{\fontfamily{cmtt}\fontseries{m}\fontshape{n}\selectfont\small{#1}}}

\newcommand{\tab}{\hspace*{1em}}

\renewcommand\footnotetextcopyrightpermission[1]{}
\begin{document}
\sloppy
\title{A Measurement Study on the (In)security of End-of-Life (EoL) Embedded Devices}



\author{Dingding Wang}
\affiliation{%
\institution{Zhejiang University}
\country{China}}

\author{Muhui Jiang}
\affiliation{%
\institution{The Hong Kong Polytechnic University}
\institution{Zhejiang University}
\country{China}}

\author{Rui Chang}
\affiliation{%
\institution{Zhejiang University}
\country{China}}

\author{Yajin Zhou}
\affiliation{%
\institution{Zhejiang University}
\country{China}}

\author{Baolei Hou}
\affiliation{%
\institution{Zhejiang University}
\country{China}}

\author{Xiapu Luo}
\affiliation{%
\institution{The Hong Kong Polytechnic University}
\country{China}}

\author{Lei Wu}
\affiliation{%
\institution{Zhejiang University}
\country{China}}

\author{Kui Ren}
\affiliation{%
\institution{Zhejiang University}
\country{China}}

\begin{abstract}

Embedded devices are becoming popular. Meanwhile, researchers are actively working on improving the security
of embedded devices.
However, previous work ignores the insecurity caused by a special category of devices, i.e., the End-of-Life (EoL in short) devices.
Once a product becomes End-of-Life, vendors tend to no longer maintain its firmware or software, including providing bug fixes and security patches.
This makes EoL devices susceptible to attacks.
For instance, a report showed that an EoL model with thousands of active devices was exploited to redirect web traffic for malicious purposes.

In this paper, we conduct the first measurement study to shed light on the (in)security of EoL devices.
To this end, our study performs two types of analysis, including the \textit{aliveness analysis} and the \textit{vulnerability analysis}.
The first one aims to detect the scale of EoL devices that are still alive.
The second one is to evaluate the vulnerabilities existing in (active) EoL devices. 

We have applied our approach to a large number of EoL models from three vendors (i.e., \code{D-Link}, \code{Tp-Link}, and \code{Netgear}) and detect the alive devices in a time period of \textit{ten months}.
Our study reveals some worrisome facts that were unknown by the community.
For instance, there exist a large number (more than $2$ million) of active EoL devices.
Some devices (nearly $300,000$) are still alive even after five years since they became EoL. 
More than half of the vulnerabilities ($182$ of $294$) are discovered after the EoL date. 
Though vendors may release security patches after the EoL date, however, the process is ad hoc and incomplete.
As a result, more than $1$ million active EoL devices are vulnerable, and nearly half of them are threatened by high-risk vulnerabilities.
Attackers can achieve a minimum of $2.79$ Tbps DDoS attack by exploiting OS command injection vulnerabilities and  compromising a large number of active EoL devices.
We believe these facts pose a clear call for more attention to deal with the security issues of EoL devices.

\end{abstract}




\maketitle
\pagestyle{plain} 

\input{intro.tex}

\input{background.tex}

\input{03_methodology.tex}

\input{04_result.tex}

\input{discussion.tex}

\input{relatedwork.tex}

\section{Conclusion}
\label{sec:conclusion}

In this paper, we conduct the first measurement study to shed light on the (in)security of EoL devices.
To this end, our study performs two types of analysis, including the \textit{aliveness analysis} and the \textit{vulnerability analysis}.
We have applied our approach to a large number of EoL models from three vendors, including \code{D-Link}, \code{Tp-Link} and \code{Netgear}.
Our study reveals some worrisome facts that were unknown by the community.
We believe these facts pose a clear call for more attention to deal with the security issues of EoL devices.

\bibliographystyle{ACM-Reference-Format}
\bibliography{ref}

\end{document}

%% file: intro.tex
\section{Introduction}

Embedded devices have become an essential part of our daily lives.
For instance, homework routers work nearly 24/7 in everyone's home
to provide network connections.
According to a report by Transforma Insights~\cite{embeddednum},
there will be 5.4 billion embedded devices connecting
to the Internet at the end of 2030.

Meanwhile, the security of embedded devices gradually
receives widespread attention.  
Once the embedded devices are attacked,
they can be abused to leak users' privacy.
Much worse, attackers can make use of a large number of compromised
devices to construct a botnet and launch the DDoS (Distributed Denial-of-Service)~\cite{ddos} attack, resulting in a serious impact on cyberspace security.
For instance, the Mirai botnet infected millions of devices and
launched over $15,000$ attacks against various targets
such as DNS service providers~\cite{antonakakis_understanding_2017}.

To this end, researchers are actively working on improving the security of embedded devices.
Their efforts can be roughly classified into two categories.
The first category is to develop new analysis techniques
to better analyze the firmware of embedded devices.
They apply software analysis techniques~\cite{shoshitaishvili_firmalice_2015,shoshitaishvili2016sok} such as the static analysis~\cite{david_firmup:_2018} and the dynamic analysis~\cite{zheng_firm-afl:_2019,davidson2013fie,chen_iotfuzzer:_2018} to locate zero-day vulnerabilities.
They also work on fingerprinting techniques to recognize
firmware components, assisting in finding N-day vulnerabilities~\cite{de_capitani_di_vimercati_towards_2017}.
The second category is to reveal the current situation
in the embedded device ecosystem based on large-scale studies,
such as weak security awareness of users or vendors~\cite{costin2014large, cui2009brave, cui2010quantitative, Bojinov, alrawi_sok:_nodate, cetin_cleaning_2019} and new kinds of attacks against embedded devices~\cite{dang_understanding_2019,he_rethinking_nodate}.

However, previous work ignores the insecurity caused by a special category of devices, i.e., the End-of-Life (EoL in short) devices.
Because of various reasons such as the shift in market demands and technology innovation, products are likely to be announced as End-of-Life (EoL) by vendors over time~\cite{dlink_eolpolicy}.
Once a product becomes End-of-Life, vendors tend to not maintain the firmware or software for it, including providing bug fixes and security patches~\cite{dlink_eolpolicy}.

Indeed, the fact that security patches are not guaranteed for EoL products results in serious security concerns.
We observe that some critical vulnerabilities exist in EoL products.
For example, five new vulnerabilities were discovered in \code{D-Link DSL-2640B} in 2020~\cite{dsl2640B}.
However, \code{D-link} announces it will not fix the vulnerabilities because \code{DSL-2640B} has become EoL since 5/5/2013.
Therefore, vulnerable EoL devices can be exploited by attackers if they are still alive on the Internet.
This happened in reality.
For example, Bad Packet reported that the malware \code{DNSChanger} compromised the active EoL model \code{D-Link DSL-2640B} which has thousands of active devices\footnote{One EoL model has multiple EoL devices, which is described in Section~\ref{sec:embedded device}} to redirect web traffics for malicious purposes~\cite{DNSattact}.
While such reports offer detailed analysis about the spotted security issues caused by EoL devices, they do not provide a systematic view of the overall situation of EoL devices.

\smallskip
\noindent \textbf{Our work}\tab
In this paper, we conduct the first measurement study to shed light on the (in)security of EoL devices, which has been overlooked.
The purpose is to reveal the current situation of EoL devices by answering the following two questions.
First, how many EoL devices are still alive and connected to the Internet?
Second, whether EoL devices have vulnerabilities? If so, whether the vulnerabilities can cause serious consequences?

To this end, our study performs two types of analysis, including the \textit{aliveness analysis} and the \textit{vulnerability analysis}.
Aliveness analysis aims to detect the scale of EoL devices that are still alive.
In particular, we leverage the cyberspace search engines for this purpose.
There exist multiple search engines with different search algorithms.
We evaluate the accuracy of them, and choose the most accurate one (i.e., \code{ZoomEye}) in our study.
Then we continuously collect information about active EoL devices.

Vulnerability analysis is to find the vulnerabilities in EoL models.
We use two methods in this study to collect the vulnerabilities.
We first use public channels, including the Common Vulnerabilities and Exposures (CVE)~\cite{cve}, National Vulnerability Database (NVD)~\cite{nvd} and Exploit Database (EDB)~\cite{edb}, to collect the information about vulnerabilities that can affect EoL models.
However, we find the information, especially the affected models of vulnerabilities, retrieved from these channels is not complete and may miss some vulnerable EoL models. 
To complement the required information, we further leverage patches released by vendors to locate more vulnerabilities.
In particular, there exist cases that new firmware images will be released to fix \textit{critical vulnerabilities} after the EoL date. 
We observed that multiple models from the same vendor share the same vulnerability while the patch is not released for every vulnerable EoL model.
Thus, 
we analyze the patches and the release notes to generate particular patterns for critical vulnerabilities.
With these patterns, we can scan the other firmware images to check whether they are vulnerable.


\smallskip
\noindent \textbf{Experiment results}\tab
We applied our approach to a large number of EoL models from three vendors, including \code{D-Link}, \code{Tp-Link} and \code{Netgear} as they publicly release the EoL models~\cite{dlink_eol_models,netgear_eol_models,tplink_eol_models}.

Specifically, our aliveness analysis started on June 23, 2020, with 287 \code{D-Link} EoL models collected from their official site.
We added two extra vendors (\code{Tp-Link} and \code{Netgear}) on November 7, 2020, and also crawled more EoL models of \code{D-Link} due to the update of its EoL model list.
In total, we collected $894$ EoL models for evaluation.
We then use \code{ZoomEye} to collect the alive EoL devices once a week from June 23, 2020 to May 10, 2021 (39 weeks).
In total, we detect $2,166,088$ active EoL devices falling in $159$ EoL models (Section~\ref{sec:aliveness}). 

We then performed the vulnerability analysis based on EoL models of \code{D-Link}.
In particular, we collect $294$ vulnerabilities for $433$ EoL models of \code{D-Link} from public channels (Section~\ref{subsubsec:public}).
These vulnerabilities affect $112$ EoL models.
Because the list of affected EoL models from NVD is incomplete, we further analyzed the security patches to locate more affected EoL models (Section~\ref{subsubsec:patch}). 
We find $13$ extra vulnerable EoL models that are missed in public channels.
In total, $125$ out of $433$ EoL models have at least one vulnerability.

Based on the experiment result, our work reveals the following worrisome facts.

\begin{itemize}[leftmargin=*]
    \item  There exist a large number (more than 2 million) of active EoL devices. Some devices (nearly $300,000$) are still alive even after five years since they became EoL, and some of them (more than $120,000$) can be reachable through insecure protocols.
    If these devices are exploited, they can cause serious consequences, as demonstrated by the Mirai botnet~\cite{antonakakis_understanding_2017}.
    
    \item  Around one third of EoL models have at least one vulnerability, and the risk of nearly half of vulnerabilities are ranked the highest.
    For instance, 33\% ($98$/$294$) vulnerabilities belong to the OS command injection. They can be used to fully control the devices.
    
    \item The vulnerable EoL models have more than 1 million alive devices, with a vulnerable rate of 92\% ($1,172,443$/$1,272,156$). More than $500,000$ active devices have high-risk vulnerabilities.
    Attackers can achieve a minimum of $2.79$ Tbps attack traffic by controlling vulnerable EoL devices with OS command injection vulnerabilities.

    \item More than half of the vulnerabilities ($182$ of $294$) are discovered after the EoL date.
    This indicates that they will not be patched in most cases. 

    \item Although vendors may release security patches after the EoL date. However, the process is ad hoc and incomplete. 
    For instance, the patches are released only for a small number of vulnerable models, and in some cases the patches do not fully function (can be bypassed).
    This shows that the vendors may not have a systematic way to track vulnerabilities in their products.
    
    \item The security patches released by vendors can provide valuable information for attackers to retrieve extra models susceptible of a vulnerability that was unknown.
    
    \item The CVE list provides valuable information of vulnerable devices, however, the information is incomplete.
    For instance, around 23\% ($68$/$294$) vulnerabilities do not have CVE numbers.
    This fact shows that the security evaluation solely based on CVE is not enough for EoL devices. 
    
\end{itemize}

\smallskip
\noindent \textbf{Contribution}\tab
In summary, this paper makes the following main contributions.

\begin{itemize}[leftmargin=*]
    \item We conduct a measurement study on the (in)security of EoL devices which has been overlooked by the community. \textit{To the best of our knowledge, this is the first systematic study on EoL devices from the security perspective}.
    \item We propose two types of analysis and apply them to EoL devices from three vendors in a period of ten months.
    \item We reveal the worrisome facts of EoL devices. We believe these results pose a clear call for more attention to the security of EoL devices.
\end{itemize}

To engage the community, we will release all the data collected in this study for further analysis.


%% file: background.tex
\section{Background}
\label{sec:background}

\subsection{Embedded Device}
\label{sec:embedded device}
Embedded devices vary a lot.
There are different kinds of embedded devices such as routers and internet cameras.
In each category of embedded devices, different kinds of vendors provide different series.
For example, the \code{DIR} series is one router series released by \code{D-Link}.
In each series, there are various models referring to the Stock Keeping Unit (SKU) of each product.
For example, \code{DIR-818LW} is one model, which belongs to series \code{DIR}. 
One model may also have some different hardware or configurations.
Thus, vendors use different sub-models to distinguish the firmware images for one model.
For example, \code{DIR-818LW\_REVA} and \code{DIR-818LW\_REVB} represent two different sub-models (or hardware configurations) of the model \code{DIR-818LW}.
For each sub-model, vendors release various versions of firmware such as \code{DIR-818LW\_REVA\_FIRMWARE\_1.01B04}.

\subsection{Cyberspace Search Engine}
Cyberspace search engines are designed to search internet-connected
devices.
They usually leverage network scanners such as \code{NMap}~\cite{nmap} to scan the entire network
and collect various information such as IP address.
They also send packets to target hosts and analyze the response packets to collect detailed information such as the model and firmware version of embedded devices.

Cyberspace search engines are used by security researchers
to analyze the impacts of malware and attackers to seek vulnerable targets.
There are several widely used cyberspace search engines: \code{ZoomEye}~\cite{zoomeye}, \code{Shodan}~\cite{shodan}, and \code{BinaryEdge}~\cite{binaryedge}.
\code{ZoomEye} uses the network scanner \code{XMap} developed based on \code{NMap}. 
\code{Shodan} is the first cyberspace search engine in the world,
used by lots of large enterprises.
\code{BinaryEdge} provides service for some security companies
to discover and analyze malware.
These search engines also provide search parameters for users to search specific devices.

\subsection{Vulnerability Metrics}
The list of Common Vulnerabilities and Exposures (CVE)~\cite{cve}
contains entries for publicly known vulnerabilities.
CVE numbers (or entries) are widely used and even become metrics to evaluate security works.
National Vulnerability Database(NVD)~\cite{nvd} is a vulnerability database built upon and fully synchronized with the CVE entries.

In addition, NVD provides enhanced information such as category
and risk rank for each vulnerability. Common Weakness Enumeration (CWE)~\cite{cwe}
is a list of software and hardware weakness types.
NVD selects part of the weakness in CWE as classification criteria
to decide the category of vulnerabilities. The criteria are updated over time.
Now NVD uses CWE-1003 classification criteria~\cite{cwe-1003}.

Common Vulnerability Scoring System (CVSS) is an open
framework for communicating the characteristics and severity of vulnerabilities~\cite{cvss} .
There are two versions of CVSS: CVSSv2 and CVSSv3.
NVD uses both CVSSv2 and CVSSv3 to calculate risk scores and decides risk ranks based on risk scores for vulnerabilities.

%% file: 03_methodology.tex
\section{Methodology}
\label{sec:methodology}

\begin{figure*}[t]
	\centering
	\includegraphics[width=0.8\textwidth]{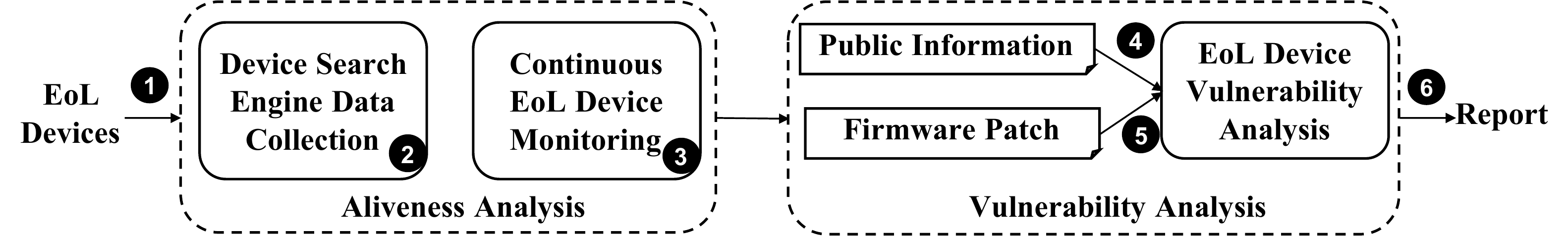}
	\caption{The flow of our study.}
	\label{fig:flow}
\end{figure*}

The purpose of this work is to perform a measurement study of vulnerable EoL devices in the wild.
Figure~\ref{fig:flow} shows the flow of our study, including the \textit{aliveness analysis} and \textit{vulnerability analysis}.
Aliveness analysis aims to reveal the number of EoL devices that are still alive, and the vulnerability analysis is to find the vulnerabilities in EoL models.
Due to the nature of EoL devices that vendors may stop the software updates, the existence of vulnerabilities and the aliveness of these devices can raise the alert on the security impact of them.

Specifically, our study first collects a list of EoL models (\ding{182}) and then leverages the public cyberspace search engines, i.e., \code{ZoomEye}, to detect the EoL devices that still connect to the Internet (\ding{183}).
We repeat this process once a week for around \textit{ten months} (\ding{184}).
For each EoL model, we also collect the vulnerabilities that affect these models from public channels, including CVE, NVD, and public exploits(\ding{185}).
Because the public information may be incomplete, i.e., some vulnerable models are not in the list of affected models from NVD, we also leverage the patches and release notes from the vendors to further locate the vulnerabilities existing in the EoL models (\ding{186}).
Finally we evaluate the security impacts of EoL devices and generate the final report (\ding{187}).

In the following, we elaborate the two types of analysis in Section~\ref{sec:methodology} and report the detailed result in Section~\ref{sec:result}.

\subsection{EoL Device Collection}
\label{sec:collection}
We take the following methods to select the EoL models used in this study.
First, the vendors should be popular to make the study representative.
Second, the vendors should maintain a publicly available list of EoL models.
For instance, \code{D-Link} maintains the information of EoL models on its website~\cite{netgear_eol_models}.
We leverage web crawlers to retrieve such information from three popular vendors, including \code{D-Link}~\cite{dlink_eol_models}, \code{Netgear}~\cite{netgear_eol_models}, and \code{Tp-Link}~\cite{tplink_eol_models}.

\subsection{Aliveness Analysis}
\label{sec:aliveness}

The aliveness analysis aims to detect active EoL devices.
If there are a large number of active EoL devices which may have vulnerabilities, they can pose serious security threats.
In this study, we use popular cyberspace search engines to collect the data in a \textit{ten-month} period to perform a continuous observation.

\subsubsection{Challenges}
There are two main challenges in getting an accurate number of active EoL devices.

{First, different cyberspace search engines can return different results as they use different detection algorithms.}
There are several popular cyberspace search engines such as \code{Shodan}~\cite{shodan}, \code{ZoomEye}~\cite{zoomeye}, and \code{BinaryEdge}~\cite{binaryedge}.
However, we observe that when searching for specific devices, they return different results.
For example, \code{BinaryEdge}, \code{Shodan}, and \code{ZoomEye} show that the EoL model \code{D-Link DCS-5009L} has $84$, $213$, and $8,486$ active devices, respectively.

{Second, the returned results of active devices from cyberspace search engines can be outdated.}
Due to the performance concern, search engines would cache the search result in the database and return the detected devices back while receiving the user request.
However, devices may not be active after a period of time but still exist in the database of search engines.
In this case, search engines can return the devices that were active a few months (or even years) ago.

\subsubsection{Our solution}
To this end, we evaluate cyberspace search engines to select the most accurate one and use it to build the database of active EoL devices of selected models in this work.
Besides evaluating the engines, we also verify the activeness of the returned results of the selected engine during each search.

\input{tables/search_engines.tex}

\smallskip
\noindent \textbf{Select search engines}\tab
Cyberspace search engines usually provide search parameters for users to filter the devices of specific models, i.e., \textit{app} in \code{ZoomEye}, \textit{product} in \code{Shodan} and \code{BinaryEdge}.
However, different engines can still return different results for the same keyword.
In this work, we evaluate the accuracy of three widely-used search engines, i.e., \code{ZoomEye}~\cite{zoomeye}, \code{Shodan}~\cite{shodan}, and \code{BinaryEdge}~\cite{binaryedge}, and then leverage the most accurate one to build the database of active EoL devices.

Specifically, we randomly select ten EoL models and search the active devices with \code{ZoomEye}, \code{Shodan}, and \code{BinaryEdge}.
We then verify the correctness of the search engine by sending specific packets.
In particular, we send the ICMP (ping) packet to the active devices reported by search engines.
If the device sends back the ICMP packet, it is marked as alive.
However, the ICMP protocol may be disabled by some devices.
Thus, if we receive no response after sending the ICMP packet, we further send the TCP SYN packet to see whether they will respond packets (e.g., TCP ACK packets or TCP RST packets), which means they are active.
We also utilize \code{NMap}~\cite{nmap} to check the models of active hosts.
For the hosts that \code{Nmap} cannot recognize their models or the recognized models from \code{Nmap} are different from cyberspace search engines, we manually analyze the packets and double-check their models.

The evaluation results are shown in Table~\ref{tab:search_engine}.
For EoL models that have more than $500$ active devices, we randomly select 500 ones.
The evaluation results shows that \code{ZoomEye} can return the most active EoL devices (the column \textit{Accurate Amount} in the table~\ref{tab:search_engine} - the product of the accuracy rate and the number of active devices returned by search engines).
Although \code{Shodan} and \code{BinaryEdge} can reach a higher accuracy rate for some specific EoL models, these two search engines cannot detect some EoL models that \code{ZoomEye} can detect.

\input{tables/zoomeye.tex}

Therefore, we collect information of active EoL devices with \code{ZoomEye} in this work.
Once we get a new EoL model, we detect active devices and retrieve the detailed information such as IP address, location (as shown in Table~\ref{tab:attribute}) of the device from \code{ZoomEye}.
We continuously collect the information of active devices for each EoL model every week to conduct a long-term observation.

\smallskip
\noindent \textbf{Verify the search results}\tab
Since the search engine can return outdated results, we verify the returned results of \code{ZoomEye} during each query.
In particular, \code{ZoomEye} provides a maximum number of $10,000$ detailed records that contain the information in Table~\ref{tab:attribute}.
Every time we get a new detailed record from \code{ZoomEye}, we extract the IP address and the port number from the record.
Because TCP and UDP are the two most popular protocols in the transport layer of the network, we first send TCP SYN packets to the IP address.
If we do not receive any response packet, we then send UDP packets again to the same destination.
We send verification packets from multiple locations, including Mainland China, the United States of America, and Hong Kong.
If the device sends back response packets, we treat it as an alive device.

\subsection{Vulnerability Analysis}
\label{sec:vulnerability}

The vulnerability analysis aims to reveal the insecurity in EoL devices based on public vulnerabilities.
As the name \textit{End-of-Life} suggests, vendors do not tend to release security patches for EoL devices.
Thus, if vulnerabilities exist in EoL devices, they may exist forever and can be exploited to launch further attacks.

In this work, we use two mechanisms to collect the vulnerabilities of EoL devices. First, we use public channels, including CVE~\cite{cve}, NVD~\cite{nvd}, and EDB~\cite{edb}, to collect the vulnerability information.
However, we find the information, especially the affected models, retrieved from these channels, is not complete.
Thus, it may miss some vulnerable EoL devices.
Second, we further leverage the patch released by the vendors to locate more vulnerabilities.
In particular, there exist cases that new firmware images are released to fix \textit{critical vulnerabilities} after the EoL date.
We can analyze the patches and the release notes together to generate particular patterns for critical vulnerabilities and then use these patterns to scan the latest firmware images~\footnote{The latest firmware denotes the last released firmware image for the EoL model, usually before the EoL date.} to check whether they are vulnerable.
That's because multiple models from the same vendor may share the same vulnerability, but the patch is not released for every vulnerable EoL model.

\subsubsection{Retrieve vulnerabilities from public channels} 
\label{subsubsec:public}
There are some challenges when retrieving the vulnerabilities from public channels.

First, {some vulnerabilities do not have CVE numbers.}
The CVE~\cite{cve} number denotes a publicly known vulnerability.
However, we observe that some vulnerabilities do not have CVE numbers.
For example, \code{D-Link} reported a critical XSS (Cross-Site Scripting) vulnerability that can be exploited in some EoL models~\cite{xsspatch}.
However, this vulnerability does not have the corresponding CVE number.

Second, the vulnerability information is inaccurate.
The NVD~\cite{nvd} is a vulnerability database that provides rich information for each vulnerability such as the CWE category, affected models, and risk ranks.
However, there are two problems with the information from NVD.
First, because NVD is fully synchronized with CVE, the vulnerabilities that are not in the CVE entries do not have this information.
Second, the CWE categories of some vulnerabilities provided by NVD are incorrect, as reported by the previous study~\cite{liu_large-scale_2020}.

\smallskip
\noindent \textbf{Our solution}\tab
First, we collect vulnerabilities from various sources other than the CVE list and NVD.
In particular, we collect exploit scripts from EDB(Exploit Database)~\cite{edb} which includes exploit scripts from various sources such as the exploit framework \code{Metasploit}~\cite{metasploit}.
Then we obtain the affected models of the vulnerabilities which are exploited in these scripts.
If the vulnerability affects EoL models and does not have a CVE number, we also count it in this work.

Second, we manually investigate whether the CWE categories
of the vulnerabilities are correct. We also manually extract
enhanced information from various sources for the vulnerabilities that
do not have CVE numbers. In particular, 
for the CWE categories, we adopt a unified classification standard
(CWE-1003 classification criteria) for all the vulnerabilities to
further analyze the type distribution.

Note that, we conduct a verification process
to show the vulnerabilities we collect really exist in EoL models
in the following three ways.
\begin{itemize}[leftmargin=*]
	\item \textbf{Dynamic analysis.} We emulate the firmware images of EoL models with \code{FIRMADYNE}~\cite{chen2016towards} and
	utilize Proof of Concept (PoC) to confirm the existence of vulnerability.
	\item \textbf{Static analysis.} We manually check whether the vulnerable function exists by
 disassembling the firmware images with \code{IDA Pro}~\cite{ida} and \code{Ghidra}~\cite{ghidra}.
	\item \textbf{Real device tests.} We collect real EoL devices whose firmware images can not be emulated by \code{FIRMADYNE} from the e-commerce website (i.e., Amazon ) and utilize PoC to confirm whether they are vulnerable.
\end{itemize}

\subsubsection{Retrieve vulnerabilities from firmware patches} 
\label{subsubsec:patch}
We observed that new firmware images are released for EoL models to fix \textit{critical vulnerabilities}.
Although this behavior alleviates the security issue, they provide valuable channels for security researchers (and attackers) to retrieve the vulnerability information. 
Based on this observation, we analyze patches to detect vulnerable models that are missed in public channels. 

\smallskip
\noindent \textbf{Step I: Retrieve firmware images}\tab
To analyze the patches released after the EoL date,
we download the firmware images from the vendors' websites
and select the images released after the EoL date.

However, due to the chaotic management of release notes, it is non-trivial to obtain the release time of the firmware images. For instance, 
not all the firmware images have release notes and not all the release notes contain the release date. Furthermore, we find the release date inside the release note may be inaccurate (e.g., earlier than build time of firmware).
Thus, we retrieve the release time of a firmware image by combining the release time inside release notes
with date information from the firmware image. 
For instance, some firmware images contain the file \code{/etc/config/builddaytime} which shows the time when the firmware is built.
The \code{/etc/sysinfo} in some images also contains the firmware build time.


\smallskip
\noindent \textbf{Step II: Analyze firmware patches}\tab
We compare the differences in the firmware released
after the EoL date with
the firmware released before the EoL date to locate the patches.
We need to confirm whether the difference in the firmware image
is to fix a vulnerability. 
To this end, we manually retrieve the release notes of the firmware images
and understand the intentions of the firmware release.
Specifically, there are two reasons for a firmware release.
The first is to provide new functionalities or features.
The second is to provide security-related patches of vulnerabilities, e.g., ``Fixed Remote Command Execution vulnerability (CVE-2019-16057)".

We analyze release notes to retrieve security-related patch information.
If the release note contains the CVE number or the clear vulnerability description,
it's easy for us to retrieve the patched function or files.
However, the release reasons sometimes are ambiguous such as
``Closed a publicly disclosed potential vulnerability" without mentioning
the vulnerability clearly. In this case, we manually analyze the different
files to locate the patched functions or files.

\smallskip
\noindent \textbf{Step III: Scan all firmware images}\tab
After Step II, we manually build the vulnerability pattern for each patched vulnerability and scan the other firmware images.
With these patterns, we can locate the existence of the vulnerabilities in the firmware images of other EoL models.

This process serves the following purpose.
In particular, it can find the EoL models that have the same vulnerability but do not have a patch.
This seems unusual, since if the vendor patched one EoL model, other affected models should also be patched.
But in reality, the vendor will not patch all the affected models.
For instance, \code{D-Link} released a security patch of CVE-2013-7471 for the model \code{DIR-815} after the EoL date but not for \code{DAP-1522}, which is discussed in detail in Section~\ref{subsec:result_vulnerability}.

%% file: tables/search_engines.tex

\begin{table*}[t]
\centering
\caption{Evaluation results of cyberspace search engines. \textit{All} means the number of active devices that search engines return. \textit{Sample} means the number of devices that we verify. \textit{Success} means the number of correct active devices that we confirm their models. \textit{Accuracy Rate} is the proportion of correct active devices (e.g., \textit{Success} / \textit{Sample}). \textit{Accurate Amount} is to estimate the number of correct active devices based on the \textit{Accuracy Rate} (e.g., \textit{All} * \textit{Accuracy Rate}). For each EoL model, we bold the most \textit{Accurate Amount}.}
\label{tab:search_engine}
\footnotesize
\resizebox{\linewidth}{!}{%
\begin{tabular}{c|ccccc|ccccc|ccccc} 
\toprule
\toprule[0.5pt]
\multirow{2}{*}{\begin{tabular}[c]{@{}c@{}} EoL\\Models \end{tabular}} & \multicolumn{5}{c|}{BinaryEdge} & \multicolumn{5}{c|}{Shodan} & \multicolumn{5}{c}{ZoomEye} \\ 
\cline{2-16}
 & All & Sample & Success & \begin{tabular}[c]{@{}c@{}}Accuracy\\Rate \end{tabular} & \begin{tabular}[c]{@{}c@{}}Accurate\\Amount \end{tabular} & All & Sample & Success & \begin{tabular}[c]{@{}c@{}}Accuracy\\Rate \end{tabular} & \begin{tabular}[c]{@{}c@{}}Accurate\\Amount \end{tabular} & All & Sample & Success & \begin{tabular}[c]{@{}c@{}}Accuracy\\Rate \end{tabular} & \begin{tabular}[c]{@{}c@{}}Accurate\\Amount \end{tabular} \\ 
\midrule
DCS-5009L & 84 & 84 & 56 & 66.67\% & 56 & 213 & 213 & 118 & 55.40\% & 118 & 8,486 & 500 & 118 & 23.60\% & \textbf{2,003}  \\ 
\rowcolor{mygray}
DI-524 & 215 & 215 & 127 & 59.07\% & 127 & 537 & 500 & 165 & 33.00\% & 177 & 7,306 & 500 & 168 & 33.60\% & \textbf{2,455}  \\ 
DIR-815 & 146 & 146 & 78 & 53.42\% & 78 & 221 & 221 & 145 & 65.61\% & 145 & 868 & 500 & 143 & 28.60\% & \textbf{248}  \\ 
\rowcolor{mygray}
DCS-935L & 2,197 & 500 & 355 & 71.00\% & 1,560 & 0 & 0 & 0 & - & 0 & 8,734 & 500 & 297 & 59.40\% & \textbf{5,188}  \\ 
DCS-931L & 192 & 192 & 106 & 55.21\% & 106 & 496 & 496 & 200 & 40.32\% & 200 & 3,521 & 500 & 142 & 28.40\% & \textbf{1,000}  \\ 
\rowcolor{mygray}
DCS-910 & 386 & 386 & 266 & 68.91\% & 266 & 71 & 71 & 43 & 60.56\% & 43 & 2,397 & 500 & 314 & 62.80\% & \textbf{1,505}  \\ 
DIR-632 & 23 & 23 & 22 & 95.65\% & 22 & 0 & 0 & 0 & - & 0 & 2,317 & 500 & 175 & 35.00\% & \textbf{811}  \\ 
\rowcolor{mygray}
DCS-2136L & 0 & 0 & 0 & - & 0 & 1 & 1 & 1 & 100.00\% & 1 & 149 & 149 & 60 & 40.27\% & \textbf{60}  \\ 
DIR-130 & 0 & 0 & 0 & - & 0 & 0 & 0 & 0 & - & 0 & 215 & 215 & 126  & 58.60\% & \textbf{126}  \\ 
\rowcolor{mygray}
DCS-5029L & 0 & 0 & 0 & - & 0 & 0 & 0 & 0 & - & 0 & 5,281 & 500 & 350 & 70.00\% & \textbf{3,697}  \\
\bottomrule[0.5pt]
\bottomrule
\end{tabular}
}
\end{table*}

%% file: tables/zoomeye.tex

\begin{table}[t]
\centering
\caption{Detailed information of active EoL devices.}
\footnotesize
\label{tab:attribute}
\resizebox{\linewidth}{!}{%
\begin{tabular}{c|c} 
\toprule
\toprule[0.5pt]
Attribute & Description \\ 
\midrule
IP & The IP address of the device \\ 
\rowcolor{mygray}
Port & The port that the device uses to send packets \\ 
Service & The service that the device uses to send packets \\ 
\rowcolor{mygray}
Country & The country that the device is in \\ 
Location & The location of the device \\ 
\rowcolor{mygray}
\begin{tabular}[c]{@{}c@{}}Autonomous System\\Number(ASN) \end{tabular} & The ASN of the device \\ 
\begin{tabular}[c]{@{}c@{}} Internet Service\\Provider(ISP) \end{tabular} & The ISP of the device \\ 
\rowcolor{mygray}
Organization & The organization that the device belongs to \\ 
Banner & The packet context \\ 
\rowcolor{mygray}
Timestamp & The time when ZoomEye scans the device \\
\bottomrule[0.5pt]
\bottomrule
\end{tabular}
}
\end{table}

%% file: 04_result.tex
\section{Experiment Results}
\label{sec:result}

We applied the previous methodology to more than 800 EoL models.
In this section, we first report the experiment setup of selected EoL models and then elaborate the results of alive EoL devices and vulnerabilities in them.

\subsection{EoL Models}

In this work, we study EoL devices from three popular vendors,
including \code{D-Link}, \code{Tp-Link}, and \code{Netgear}.
They maintain web-pages for EoL model information\cite{dlink_eol_models,netgear_eol_models,tplink_eol_models}.

\begin{figure}[t]
	\centering
	\includegraphics[width=0.47\textwidth]{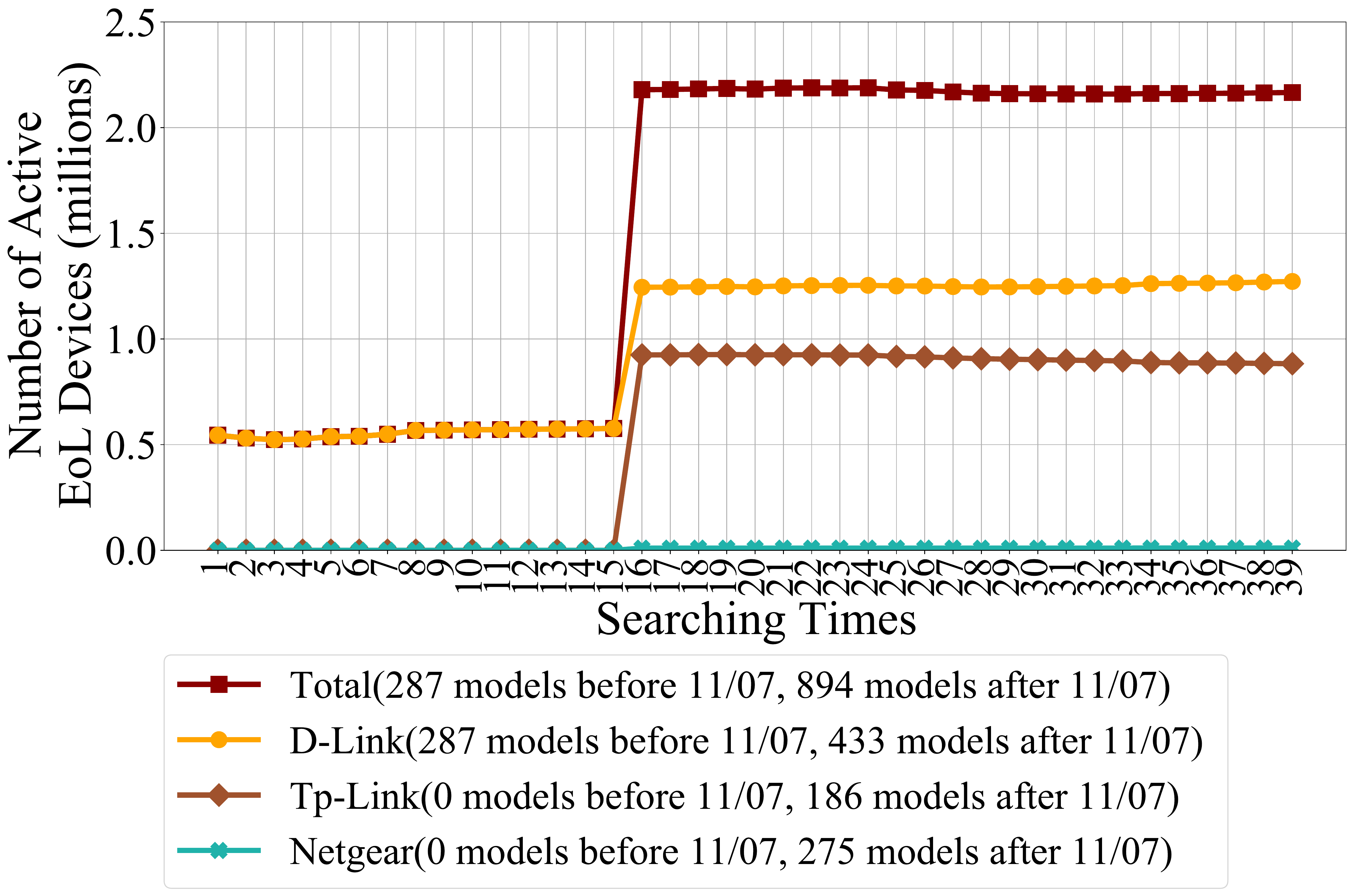}
	\caption{The number of alive EoL devices monitored in 39 weeks. The monitor begins on 6/23/2020 and stops on 5/10/2021. The $16^{th}$ time is on 11/07/2020.}
	\label{fig:actives}
\end{figure}

We start our study on June 23, 2020, with 287
\code{D-Link} EoL models collected from its website.
Then on November 7, 2020, we add two extra
vendors (\code{Tp-Link} and \code{Netgear}) and also crawl
more EoL models of \code{D-Link} because of the update of its EoL model list.
In total, we collect $894$ EoL models from three vendors.
Figure~\ref{fig:actives} shows the EoL model distribution of each vendor.

\subsection{Aliveness Analysis}
\label{subsec:ali_result}
For each EoL model, we query the alive devices using the cyberspace search engine \code{ZoomEye} once a week (Section~\ref{sec:aliveness}), from June 23, 2020 to May 10, 2021 (39 weeks).
Figure~\ref{fig:actives} shows the overall data.
The x-axis shows the index of the querying (once a week) and the y-axis shows the total number of active EoL devices.
Note that, because we added two extra vendors and more EoL models on November 7, 2020, there is a boost of the active EoL devices in the figure.

In total, we detect $2,166,088$ active EoL devices of $159$ active EoL models in the latest search (i.e., May 10, 2021).
We further analyze the data and illustrate our findings in the following.  

\smallskip
\noindent \textbf{The number of active EoL devices is stable.}\tab
Instead of decreasing steadily (that one may expect),  the amount of active EoL devices is stable, unless the number of EoL models changes.
One potential reason is that while vendors may stop supporting EoL devices, they can still be purchased from the market.
For instance, $40$ EoL models of \code{D-Link} are still sold on Amazon. 

\begin{figure}[t]
	\centering
	\includegraphics[width=0.47\textwidth]{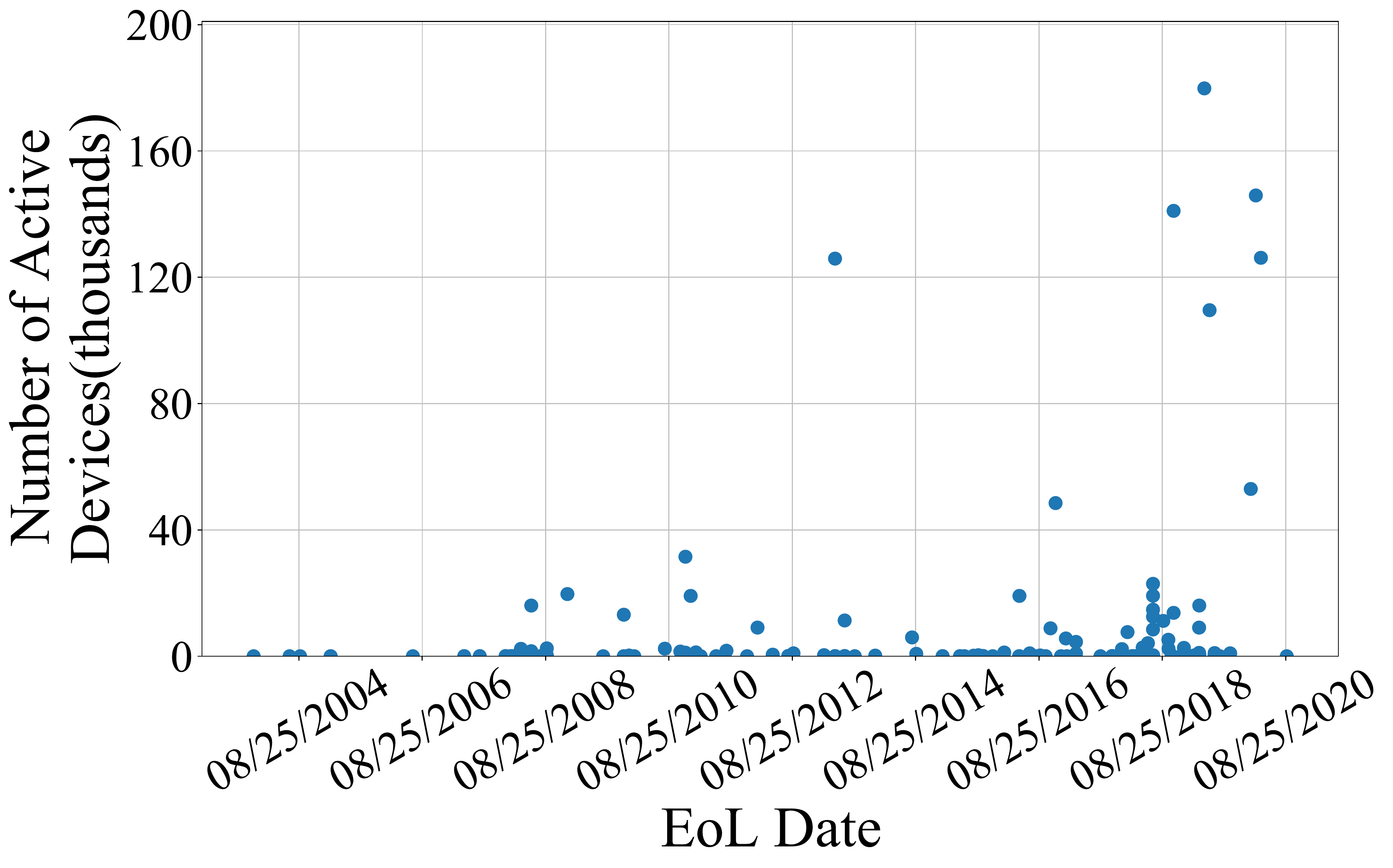}
	\caption{The correlation between the number of active devices and the EoL date of active models. The EoL date is the date of the last day of support announced by vendors.}
	\label{fig:active_date}
\end{figure}

\smallskip
\noindent \textbf{EoL devices can be active for many years after the EoL date.}\tab
Figure~\ref{fig:active_date} shows the number of active EoL devices (the y-axis) and the EoL date\footnote{EoL models of \code{Tp-Link} do not have the EoL date.} (the x-axis) of $135$ EoL models that have active devices.
Most active models are EoL within three years.
However, we also observe that $73$ (54\%) EoL models are still alive on the Internet even though they have been EoL for more than five years. 
These $73$ EoL models have a total of $293,478$ active devices.
Among the $73$ EoL models, $19$ EoL models have more than $1,000$ active devices. 
For example, \code{DSL-2640B}, the model with $125,907$ active devices, has been EoL for more than seven years.

\input{tables/type}

\smallskip
\noindent \textbf{Alive EoL devices are using insecure protocols.}\tab
Telnet and UPnP protocols have been exploited to launch further attacks due to their insufficient authentication~\cite{antonakakis_understanding_2017,upnp}.
For instance, the \code{Mirai} botnet was reported to exploit open Telnet ports~\cite{antonakakis_understanding_2017}.
However, we still find that $128,759$ and $7,607$ active EoL devices are reachable through the Telnet and UPnP protocol, respectively.

\begin{figure}[t]
	\centering
	\includegraphics[width=0.47\textwidth]{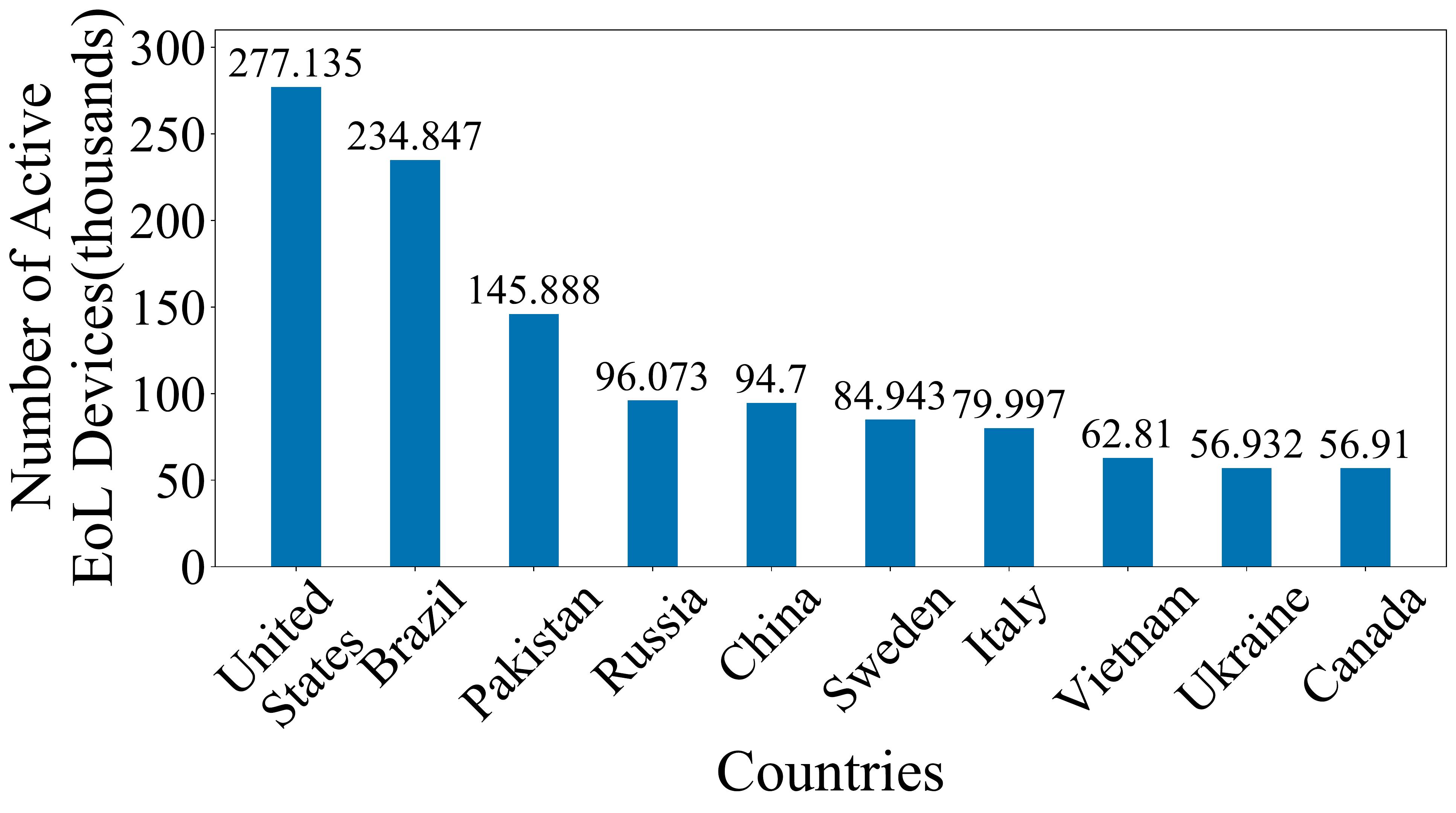}
	\caption{Alive EoL devices in top 10 countries.}
	\label{fig:country}
\end{figure}

\smallskip
\noindent \textbf{The distribution of alive EoL devices is unbalanced.}\tab
First, the type of active EoL devices is unbalanced, as shown in Table~\ref{tab:type}.
In particular, the router has the most active devices followed by internet cameras.
Second, the vendor distribution is unbalanced.
\code{D-Link} and \code{Tp-Link} have more active EoL devices than \code{Netgear}.
One potential reason is that EoL models of \code{Netgear} do not belong to the popular type (routers and internet cameras), resulting in a small number of active devices.
Third, the geographical distribution is unbalanced.
As shown in Figure~\ref{fig:country}, the United States has the most active EoL devices followed by Brazil and Pakistan.

\begin{framed}
	\vspace{-0.1in}
		\noindent
		\textbf{Summary}\tab
		The fact that there exist a large number of active EoL devices for a long time and they can be reachable through insecure protocols can bring security threats.
		If these devices are vulnerable and exploited, they can cause serious consequences. 
		\vspace{-0.1in}
\end{framed}

\subsection{Vulnerability Analysis}
\label{subsec:result_vulnerability}

We focus on EoL models of \code{D-Link} to perform the vulnerability analysis because \code{D-Link} provides EoL models for popular categories with complete information.
In total, we collect $294$ vulnerabilities for EoL models of \code{D-Link} (Section~\ref{subsubsec:public}).
These vulnerabilities affect $125$ models out of $433$ EoL models of \code{D-Link}.
In particular, $89$ EoL models have high-risk vulnerabilities.


Note that, because the list of affected models from NVD is incomplete, we analyze the security patches to locate more affected EoL models for measuring the insecurity more accurately, which find $13$ vulnerable EoL models that are missed from public channels.
In particular, we analyze the differences of the firmware image pair before and after the EoL date (Section~\ref{subsubsec:patch}), and confirm that they are security patches for $12$ vulnerabilities (Table~\ref{tab:vulnerabilities}).
We then use the pattern of these vulnerabilities to scan all the firmware images.

To analyze patches and locate more vulnerable EoL models, we collect $1,247$ firmware images for $277$ models which have firmware images among the $433$ EoL models of \code{D-Link}.
Among the $1,247$ firmware images, $686$ firmware images can be extracted and the following analysis is based on these $686$ firmware images which involve $149$ EoL models.
Among the $686$ firmware images, $36$ firmware images that involve $27$ EoL models are released after the EoL date.
That means the vendor has released new firmware images (or patches) for $27$ models even after the EoL date.

We summarize our findings in the following.

\input{tables/patch}

\smallskip
\noindent \textbf{Around one third EoL models have at least one vulnerability.}\tab
In total, $125$ EoL models have at least one vulnerability.
Among the $125$ EoL models, $112$ are detected through public channels (Section~\ref{subsubsec:public}) and extra $13$ are detected through firmware patches (Section~\ref{subsubsec:patch}).
With patches, we detect $49$ vulnerable EoL models in total, $36$ of them are already found through public channels.
Besides the extra $13$ vulnerable models, we also find more vulnerabilities in the models detected through public channels.
In particular, the risk of $6$ models are increased by newly found vulnerabilities with firmware patches.
For example, NVD shows that the model \code{DIR-412} only has two medium-risk vulnerabilities, but we find three high-risk vulnerabilities in this model through firmware patches.

\begin{figure}[t]
	\centering
	\includegraphics[width=0.47\textwidth]{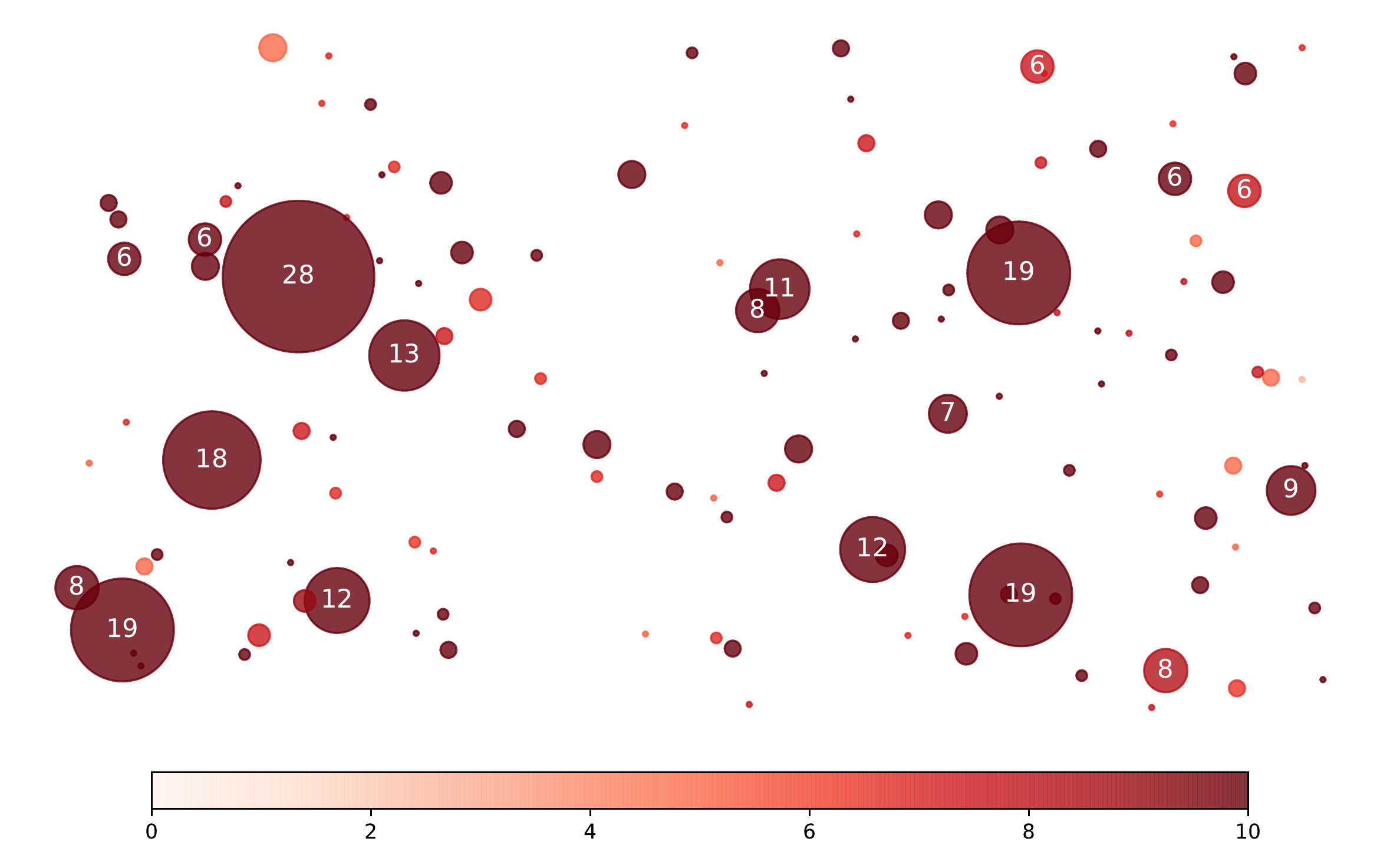}
	\caption{The heat map of vulnerable EoL models. Each node denotes a vulnerable EoL model, and the size (and the number inside the node) denotes the number of vulnerabilities. The color denotes the severity of the vulnerabilities. 
	The rank score 10 denotes the most severe vulnerability.}
	\label{fig:vul_model}
\end{figure}

Figure~\ref{fig:vul_model} shows the heat map of vulnerable EoL models.
Each node in the graph represents a vulnerable EoL model, and the size denotes the number of vulnerabilities detected in that model.
The color denotes the severity of the vulnerability (measured by the CVSSv2 framework~\cite{cvss}).
In particular, $89$ EoL models have high-risk vulnerabilities.
Moreover, $87$ models have more than one vulnerability, and $9$ of them even have more than $10$ vulnerabilities.
The \code{DIR-850L} model has the highest number of vulnerabilities (i.e., 28) with $11,190$ active devices.

\smallskip
\noindent \textbf{More than half of the vulnerabilities are discovered after the EoL date.}\tab
This implies that their affected models have a higher possibility of not being patched.
For example, \code{DIR-600} became End-of-Life on 12/1/2010, but 16 vulnerabilities in this model were discovered from 2013 to 2019.

In total, out of $294$ vulnerabilities, $182$ of them ($62\%$) were discovered after the EoL date ($141$ vulnerabilities have CVE entries, and $41$ do not.)
These vulnerabilities affect $68$ EoL models.
Among these EoL models, $33$ have active devices, with a total of $673,511$ active devices.
The most active EoL model \code{DCS-932L} has $179,806$ active devices and has been EoL for two years.
Moreover, according to CVSSv2~\cite{cvss}, $61$ vulnerabilities discovered after the EoL date are ranked as high-risk, leading to potentially serious consequences, e.g., completely being controlled by attackers.

\begin{figure}[t]
	\centering
	\includegraphics[width=0.47\textwidth]{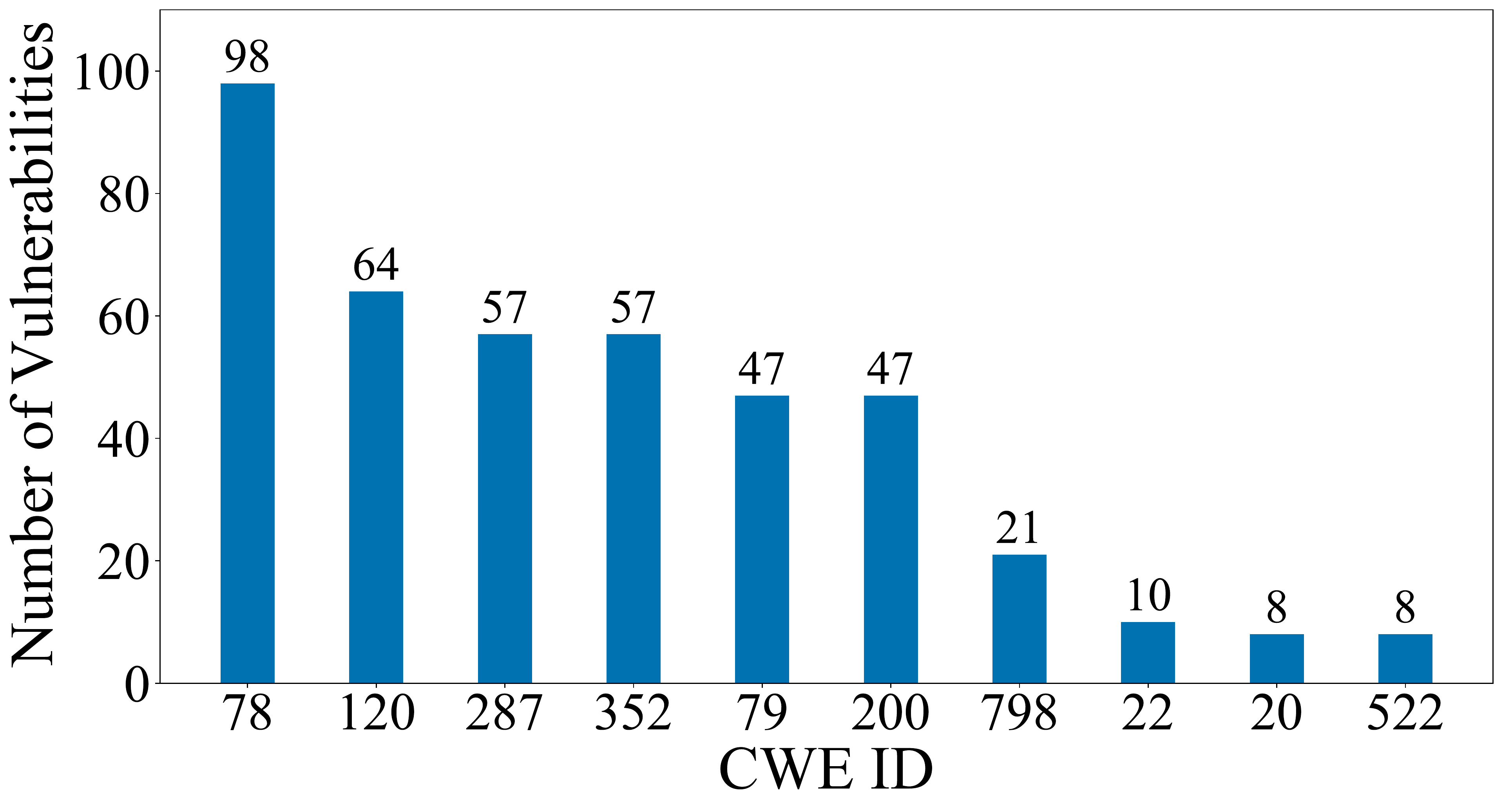}
	\caption{Top 10 types of vulnerabilities. (CWE-78: OS Command Injection; CWE-120: Classic Buffer Overflow; CWE-287: Improper Authentication; CWE-352: Cross-Site Request Forgery; CWE-79: Cross-site Scripting; CWE-200: Exposure of Sensitive Information to an Unauthorized Actor; CWE-798: Use of Hard-coded Credentials; CWE-22: Path Traversal; CWE-20: Improper Input Validation; CWE-522: Insufficiently Protected Credentials)}
	\label{fig:Type}
\end{figure}

\smallskip
\noindent \textbf{Nearly half of vulnerabilities of EoL models are high-risk.}\tab
Following the CWE-1003 classification criteria used by NVD, we calculate the distribution of vulnerability types.
As shown in Figure~\ref{fig:Type}, the \textit{OS command injection} is the most common vulnerability type, followed by buffer overflow.
The OS command injection vulnerability can be easily exploited to execute arbitrary commands and fully control the device. 
In addition, three of the top five types of vulnerabilities are associated with the web servers that can be remotely (or locally) exploited through the HTTP protocol.

\input{tables/rank}

Nearly half of the vulnerabilities are ranked as the highest risk level.
Table~\ref{tab:rank} shows the risk rank distribution of the vulnerabilities that are in CVE entries. 
Specifically, for $226$ vulnerabilities that have CVSSv2 scores, $97$ of them are ranked as high-risk.
Among the $177$ vulnerabilities with CVSSv3 scores~\footnote{Some vulnerabilities do not have CVSSv3 rank because they are discovered before the invention of CVSSv3.}, $68$ of them are ranked as critical-risk, and $79$ of them are ranked as high-risk.
It is worth noting that the vulnerabilities not in CVE entries also can cause widespread effects and rank critical.

\smallskip
\noindent \textbf{Solely using CVE to track vulnerabilities of EoL models is not enough.}\tab
Among the $294$ vulnerabilities, only $77\%$ ($226$/$294$) of them are in CVE entries.
The other $68$ vulnerabilities are located through public exploits, most of which are high-risk vulnerabilities such as remote command execution that can be exploited to control the device.
Given the fact that some security evaluation only based on CVE is not enough for EoL devices.

\input{tables/scan}


\smallskip
\noindent \textbf{Vendors may patch vulnerabilities after the EoL date. However, the patch is ad hoc and incomplete, which may expose vulnerable models that were not publicly known.}\tab
First, the patch is ad hoc. Vendors do not have a systematic way to track critical vulnerabilities and patch them.
For instance, Table~\ref{tab:scan} shows six vulnerabilities
that have security patches after the EoL date.
However, only 29\% ($28$/$97$) of vulnerable models are patched successfully.
The vendors do not systemically patch different models
or even similar models.
For example, \code{REV.A} and \code{REB.B} of
\code{DIR-815} both have the vulnerability (CVE-2019-18852).
However, only \code{REV.A} is patched and in the NVD list of affected models of this vulnerability.

Second, the security patches are not always fully functional.
For instance, the vendor has released two different patches for CVE-2018-10106.
However, one of the patches is not complete, and the vulnerability can still be exploited.
In particular, $21$ models are vulnerable, and all of them have been patched.
However, only $4$ of them have applied the right patch, while the left $17$ models have applied the incomplete patch~\cite{ineffective}.
This further confirms the ad hoc way of the vendor to apply security patches.

Third, there are cases that the previously fixed vulnerabilities appear again in later firmware images. 
For example, the firmware of \code{DAP-1522 REV.B} (version \code{2.07B01}) released in 2013 patched CVE-2019-18852.
However, the later firmware released in 2015 (version \code{PATCH\_2.03.B01}) does not apply the security patch of CVE-2019-18852.
This indicates that the vendor does not enforce a managed process to release security patches.

To make matters worse, by analyzing the released patches, attackers can locate vulnerable EoL models that are not publicly known and most likely not patched.
As shown in Table~\ref{tab:scan}, only 16\% ($16$/$97$) of vulnerable models are in the list of affected devices from NVD.
Moreover, the models that are not in the list have a lower probability of being patched.
Only 21\% ($17$/$81$) of the vulnerable models not in the known affected list are patched successfully while 69\% ($11$/$16$) of the vulnerable models in the known affected list are patched successfully.

\begin{framed}
	\vspace{-0.1in}
	\noindent
	\textbf{Summary}\tab
	The following facts further pose serious security concerns. \textit{First}, around
	one third of EoL models have at least one vulnerability, and most vulnerable EoL models have high-risk vulnerabilities.
	\textit{Second}, more than half of vulnerabilities are discovered after the EoL date, which indicates that they will not be patched in most cases.
	\textit{Third}, the ad hoc way to release security patches makes the patches incomplete and can further expose vulnerable EoL models that were not publicly known.
	\vspace{-0.1in}
\end{framed}

\subsection{Insecure EoL Devices in the Wild}
\label{subsec:measurement}

To evaluate the insecurity of EoL devices better, we first combine the results above (Section~\ref{subsec:ali_result} and Section~\ref{subsec:result_vulnerability}) and then simulate the attacks that vulnerable EoL devices can launch.
In particular, we detect $1,172,443$ vulnerable EoL devices in the latest search (i.e., May 10, 2021), as shown in Table~\ref{tab:vul_type}.
Moreover, as shown in Table~\ref{tab:attack}, the attackers can launch a $2.79$ Tbps DDoS attack, which is more than two times Mirai's attack on French web host OVH~\cite{mirai_wiki}, according to our simulation results.


\input{tables/attack_traffic.tex}

\input{tables/vul_type}

\begin{figure}[t]
\centering
\includegraphics[width=0.47\textwidth]{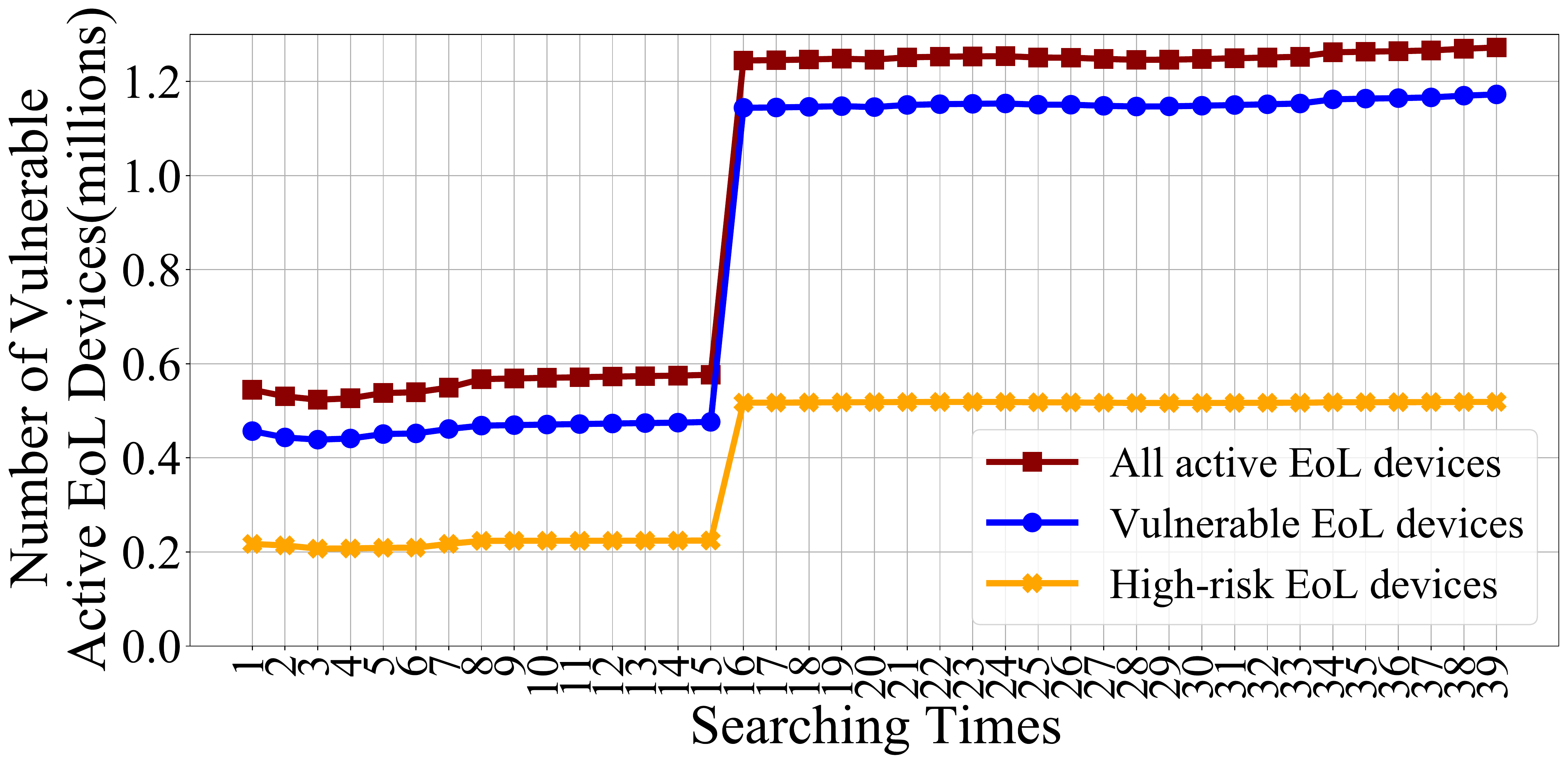}
\caption{Aliveness of Vulnerable EoL devices. The monitor begins on 6/23/2020 and stops on 5/10/2021. The boost is because we added more EoL models between $15^{th}$ and $16^{th}$ search.}
\label{fig:vul_devices}
\end{figure}

\smallskip
\noindent \textbf{The vulnerable EoL models have more than $1,000,000$ alive devices.}\tab
Out of $1,272,156$ active \code{D-Link} devices, 92\% devices (i.e., $1,172,443$) are vulnerable.
Among the vulnerable EoL devices, $518,936$ (nearly half) have high-risk vulnerabilities according to the CVSSv2 framework.
As shown in Figure~\ref{fig:vul_devices}, the amount of vulnerable active EoL devices is stable, which is the same as the amount of active EoL devices.

Moreover, Table~\ref{tab:vul_type} shows the type distribution of vulnerable active devices.
Routers and internet cameras account for more than 99\% of vulnerable EoL devices.
For instance, the \code{DSL-2640B} model has $10$ vulnerabilities with $125,907$ alive devices.
The devices have been reported to be exploited and abused by the DNSChanger malware~\cite{DNSattact}.
Five of its vulnerabilities were reported in 2020, but have not been patched yet.
In addition, the proportion of vulnerable devices in active devices is larger than the proportion of vulnerable models in active models, for routers, internet cameras, and access points.
For example, for routers, among $43$ active models $29$ ($67\%$) are vulnerable, while among $246,184$ active devices $224,782$ ($91\%$) are vulnerable.

\smallskip
\noindent \textbf{The attackers can achieve a minimum of $2.79$ Tbps attack by exploiting EoL devices.}\tab
When simulating, we only consider the OS command injection vulnerabilities and assume all devices run the latest firmware version to get a low-bound of simulated traffic.
Table~\ref{tab:attack} lists the EoL models affected by these OS command injection vulnerabilities and the number of their active devices.
There are $212,403$ active devices threatened by OS command injection vulnerabilities.
We randomly select around 15\% of their active devices to detect the bandwidth they can achieve.
We utilize the tool \code{Pathneck}~\cite{hu2004locating} to probe the bandwidth in the connection between the active devices and our server.
In total, we successfully detect bandwidth of $8,636$ active devices.
If attackers control these $8,636$ vulnerable active devices, they can achieve a $2.79$ Tbps DDoS attack.

\subsection{Case Study}
\label{subsec:case}
At last, we use \code{DIR-818LW} as a case study to demonstrate the insecurity caused by EoL devices. 
The EoL date of this model is 05/01/2017.
In 2020, we bought a \code{DIR-818LW} device
from Amazon to analyze the vulnerabilities in this model.
Note that this model has two sub-models, i.e., \code{REVA} and \code{REVB}.
The device we purchased is \code{REVA}. The original firmware version on the device is \code{DIR-818LW\_REVA\_FIRMWARE\_1.01}.
There are six firmware images for \code{DIR-818LW(REVA)}, as shown in Table~\ref{tab:818lw}.
The last firmware (\code{PATCH\_2.06}) is released after the EoL date.

\input{tables/dir-818lw.tex}

\smallskip
\noindent \textbf{Security patches that cannot be applied.}\tab
Two firmware images downloaded cannot be applied to our device,
including \code{PATCH\_2.05} and \code{PATCH\_2.06}.
Our further analysis shows that the MD5 value of the firmware
image \code{DIR-818LW\_REVA\_FIRMWARE\_PATCH\_2.05} is the same as
the firmware image \code{DIR-818LW\_REVB\_FIRMWARE\_PATCH\_2.05} which is for another sub-model (i.e., \code{DIR-818LW\_REVB}).
We suspect that the vendor released a wrong security patch.

\begin{figure}
\lstdefinestyle{mystyle}{
	commentstyle=\color{codegreen},
    keywordstyle=\color{magenta},
    stringstyle=\color{codepurple},    
	basicstyle=\footnotesize,
    breakatwhitespace=false,         
	breaklines=true,                 
    captionpos=b,                    
    keepspaces=true,                 
    numbers=left,                    
    numbersep=5pt,                  
    showspaces=false,                
    showstringspaces=false,
    showtabs=false,
	frame = single,
    tabsize=2
}
\begin{lstlisting}[language=php,style=mystyle]
$GETCFG_SVC = cut($_POST["SERVICES"], $SERVICE_INDEX, ",");
TRACE_debug("GETCFG: serivce[".$SERVICE_INDEX."] = ".$GETCFG_SVC);
if ($GETCFG_SVC!=""){
    $file = "/htdocs/webinc/getcfg/".$GETCFG_SVC.".xml.php";
    /* GETCFG_SVC will be passed to the child process. */
    if (isfile($file)=="1") dophp("load", $file);}
\end{lstlisting}
\caption{Vulnerable code in \code{/htdocs/web/getcfg.php}. The value of \$file, which can be controlled by attackers, is not checked. In this case, attackers can execute designated php script via function \textit{dophp}. }
\label{fig:cve-2018-10106}
\end{figure}

\smallskip
\noindent \textbf{Ineffective security patches.}\tab
For instance, CVE-2018-10106 leads to permission bypass and information disclosure in \code{/htdocs/web/getcfg.php}, as shown in Figure~\ref{fig:cve-2018-10106}.
The webserver gets the \code{SERVICES} parameter from received packets (Line 1) and splices it into a file name (Line 4) and executes the file (Line 6) without checking the context of the \code{SERVICES} parameter.
If an attacker sets \code{SERVICES=DEVICE.ACCOUNT}, he can directly execute \code{/htdocs/webinc/getcfg/DEVICE.ACCOUNT.xml.php} which leaks sensitive information such as username and password.

The security patch in the firmware \code{DIR-818LW\_REVA\_FIRMWARE\_1.05} adds code \textit{if(is\_power\_user() == 1)} before getting parameter in Line 1 to conduct the permission check.
In function \code{is\_power\_user()}, the web server checks whether the value of \code{AUTHORIZED\_GROUP} is empty or less than 0, which means the user is not authorized.
If so, the function \code{is\_power\_user()} returns 0 and the vulnerable code in Figure~\ref{fig:cve-2018-10106} will not be executed.
However, the permission check can be bypassed~\cite{ineffective}, making the security patch ineffective.
When handling request packets, the web server stores parameters in a string using \code{$\textbackslash{}n$} as the delimiter.
If an attacker adds \textit{$\textbackslash{}n$AUTHORIZED\_GROUP=0} to his packets, he can set \code{AUTHORIZED\_GROUP} to 0 and bypass the permission check.

 \input{tables/vul_818.tex}

\smallskip
\noindent \textbf{The public vulnerability disclosure is incomplete.}\tab
There are six vulnerabilities in
\code{DIR-818LW\_REVA\_FIRMWARE\_1.05}. All of them are
manually confirmed with the real device and exploits.
Table~\ref{tab:vul_818lw} shows the detailed information. 
However, the model \code{DIR-818LW} is not in the affected models list of all these six vulnerabilities, while there do exist public exploits that target this model.
This makes the security evaluation based on CVE ineffective.

%
%
%
%

%% file: tables/type.tex


\begin{table}[t]
    \centering
    \caption{Type distribution of active EoL devices.}
    \label{tab:type}
    \footnotesize
    \resizebox{0.99\linewidth}{!}{%
    \begin{tabular}{c|ccccccc} 
    \toprule
    \toprule[0.5pt]
    Type & Router & \begin{tabular}[c]{@{}c@{}}Internet\\Camera \end{tabular} & Gateway & \begin{tabular}[c]{@{}c@{}}Access\\Point \end{tabular} & \begin{tabular}[c]{@{}c@{}}Print\\Server~\end{tabular} & Others & \textbf{Total} \\ 
    \midrule
    \begin{tabular}[c]{@{}c@{}} EoL\\Models \end{tabular} & 164 & 99 & 183 & 118 & 13 & 317 & \textbf{894} \\ 
    \hline
    \begin{tabular}[c]{@{}c@{}} Active\\Models \end{tabular} & 60 & 44 & 32 & 14 & 7 & 2 & \textbf{159} \\ 
    \hline
    \begin{tabular}[c]{@{}c@{}} Active\\Devices \end{tabular} & 1,108,846 & 983,833 & 68,275 & 4,389 & 742 & 3 & \textbf{2,166,088} \\
    \bottomrule[0.5pt]
    \bottomrule
    \end{tabular}
    }
    \end{table}
    

%% file: tables/patch.tex

\begin{table}[t]
\centering
\caption{Vulnerabilities and the number of firmware images patching the vulnerabilities after EoL. Because KRACK is a serious and widespread attack, involving ten different CVE vulnerabilities, NVD does not provide risk rank for KRACK.}
\label{tab:vulnerabilities}
\footnotesize
\resizebox{0.75\linewidth}{!}{%
\begin{tabular}{ccc}
\toprule
\toprule[0.5pt]
Vulnerability & Risk Rank & \begin{tabular}[c]{@{}c@{}}Patched\\Firmware \end{tabular} \\ 
\midrule
CVE-2019-18852 & High & 1 \\ 
\rowcolor{mygray}
CVE-2018-6530 & High & 1 \\ 
CVE-2018-20114 & High & 3 \\ 
\rowcolor{mygray}
CVE-2014-8361 & High & 1 \\ 
CVE-2019-13561 & High & 1 \\ 
\rowcolor{mygray}
CVE-2017-7852 & Medium & 5 \\
CVE-2019-10999 & Medium & 5 \\ 
\rowcolor{mygray}
CVE-2019-16057 & High & 1 \\ 
KRACK (10 CVEs)~\tablefootnote{Key Reinstallation Attack (KRACK) exploits the vulnerability in WPA2 and involves 10 CVEs. (CVE-2017-13077, CVE-2017-13078, CVE-2017-13079, CVE-2017-13080, CVE-2017-13081, CVE-2017-13082, CVE-2017-13084, CVE-2017-13086, CVE-2017-13087, CVE-2017-13088)} & - & 6 \\ 
\rowcolor{mygray}
CVE-2020-15895 & Medium & 1 \\ 
\begin{tabular}[c]{@{}c@{}}CVE-2018-10106\\(4 CVEs)\tablefootnote{CVE-2019-17506, CVE-2020-9376, and CVE-2020-15894 also describe the same vulnerability but list different affected devices. None of these affected devices include DIR-818LW. So we use CVE-2018-10106 to stand for the vulnerability.} \end{tabular} & High & 2 \\
\rowcolor{mygray}
CVE-2013-7471 & High & 1 \\
\bottomrule[0.5pt]
\bottomrule
\end{tabular}
}
\end{table}


%% file: tables/rank.tex

\begin{table}[t]
    \centering
    \caption{The risk rank of vulnerabilities. (CVSSv2 does not have the critical risk rank.)}
    \footnotesize
    \label{tab:rank}
    \resizebox{0.35\textwidth}{!}{
        \begin{tabular}{ccc}
        \toprule
        \toprule[0.5pt]
       Risk Rank & Count (CVSSv2) & Count (CVSSv3) \\
        \midrule
        Low & 18 & 0 \\
        \rowcolor{mygray}
        Medium & 111 & 30 \\
        High & \textbf{97} & \textbf{79} \\
        \rowcolor{mygray}
        Critical & - & \textbf{68} \\
        Null & 0 & 49 \\
        \bottomrule[0.5pt]
        \bottomrule
        \end{tabular}
    }
\end{table}

%% file: tables/scan.tex

\begin{table}
\centering
\caption{The scan results of six vulnerabilities. \textit{Vulnerable Models} means the number of models that are vulnerable to the specific vulnerability. \textit{NVD List} means the number of the vulnerable models that are also in the list of affected devices from the NVD. \textit{Successful Patches} and \textit{Unsuccessful Patches} means the number of vulnerable models that are patched successfully and unsuccessfully, respectively. The number in brackets are the number of vulnerable models which are patched (successfully or unsuccessfully) and also in the list of affected devices of the NVD.}
\footnotesize
\label{tab:scan}
\resizebox{0.95\linewidth}{!}{%
\begin{tabular}{c|cccc} 
\toprule
\toprule[0.5pt]
Vulnerability & \begin{tabular}[c]{@{}c@{}}Vulnerable \\Models \end{tabular} & \begin{tabular}[c]{@{}c@{}}NVD\\List \end{tabular} & \begin{tabular}[c]{@{}c@{}}Successful\\Patches \end{tabular} & \begin{tabular}[c]{@{}c@{}}Unsuccessful\\Patches \end{tabular} \\ 
\midrule
CVE-2019-18852 & 31 & 4 & 2 (1) & - \\ 
\rowcolor{mygray}
CVE-2020-15895 & 10 & 1 & 3 (0) & - \\ 
CVE-2018-10106 & 21 & 1 & 4 (0) & 17 (1) \\ 
\rowcolor{mygray}
CVE-2013-7471 & 16 & 1 & 6 (1) & - \\ 
CVE-2017-7852 & 15 & 8 & 11 (8) & - \\ 
\rowcolor{mygray}
CVE-2019-13561 & 4 & 1 & 2 (1) & - \\ 
\textbf{Total} & \textbf{97} & \textbf{16} & \textbf{28 (11)} & \textbf{17 (1)} \\
\bottomrule[0.5pt]
\bottomrule
\end{tabular}
}
\end{table}


%% file: tables/attack_traffic.tex

\begin{table*}[t]
\centering
\caption{The result of attack simulation. \textit{Active Devices} means the number of active devices detected by \code{ZoomEye} in the latest searching (i.e., May 10, 2021). \textit{Measured Devices} means the number of devices we select randomly to measure the bandwidth. \textit{Successfully Measured Devices} means the number of devices we measure the bandwidth successfully. \textit{Average Bandwidth} and \textit{Attack} mean the average and summation  of bandwidths for each EoL model, respectively. The summation of bandwidths is the traffic that attackers can achieve in DDoS attacks.}
\label{tab:attack}
\footnotesize
\resizebox{0.9\linewidth}{!}{%
\begin{tabular}{c|ccccc} 
\toprule
\toprule[0.5pt]
EoL Model & Active Devices & Measured Devices & Successfully Measured Devices & Average Bandwidth (Mbps) & Attack (Gbps) \\ 
\midrule
DIR-412 & 250 & 89 & 25 & 291.0 & 7.1 \\ 
\rowcolor{mygray}
DIR-600 & 31,507 & 11,564 & 3,211 & 331.5 & 1,039.6 \\ 
DIR-645 & 1,434 & 459 & 86 & 519.5 & 43.6 \\ 
\rowcolor{mygray}
DIR-655 & 4,137 & 1,617 & 376 & 344.1 & 126.4 \\ 
DIR-815 & 2,258 & 873 & 172 & 398.9 & 67.0 \\ 
\rowcolor{mygray}
DIR-825 & 187 & 97 & 19 & 305.0 & 5.7 \\ 
DIR-860L & 2,749 & 1,025 & 257 & 386.9 & 97.1 \\ 
\rowcolor{mygray}
DIR-865L & 1,182 & 381 & 31 & 358.4 & 10.9 \\ 
DCS-930L & 141,027 & 10,397 & 4,056 & 329.7 & 1,305.9 \\ 
\rowcolor{mygray}
DIR-615 & 7,641 & 715 & 163 & 322.0 & 51.3 \\ 
DIR-850L & 11,190 & 596 & 158 & 498.4 & 76.9 \\ 
\rowcolor{mygray}
DIR-868L & 8,841 & 563 & 82 & 363.2 & 29.1 \\ 
\textbf{Total} & \textbf{212,403} & \textbf{28,376} & \textbf{8,636} & - & \textbf{2,860.4} \\
\bottomrule[0.5pt]
\bottomrule
\end{tabular}
}
\end{table*}

%% file: tables/vul_type.tex
\begin{table}
\centering
\caption{Type distribution of \code{D-Link} vulnerable EoL devices.}
\label{tab:vul_type}
\footnotesize
\resizebox{0.99\linewidth}{!}{%
\begin{tabular}{c|ccccccc} 
\toprule
\toprule[0.5pt]
Type & Router & \begin{tabular}[c]{@{}c@{}}Internet\\Camera\end{tabular} & Gateway & \begin{tabular}[c]{@{}c@{}}Access\\Point\end{tabular} & \begin{tabular}[c]{@{}c@{}}Print\\Server\end{tabular} & Others & \textbf{Total} \\ 
\midrule
\begin{tabular}[c]{@{}c@{}}EoL\\Models\end{tabular} & 111 & 84 & 44 & 54 & 13 & 127 & \textbf{433} \\ 
\hline
\begin{tabular}[c]{@{}c@{}}Active\\Models\end{tabular} & 43 & 43 & 5 & 11 & 7 & 2 & \textbf{111} \\ 
\hline
\begin{tabular}[c]{@{}c@{}}Vulnerable\\Models\end{tabular} & 60 & 36 & 0 & 16 & 1 & 12 & \textbf{125} \\ 
\hline
\rowcolor{mygray}
\begin{tabular}[c]{@{}c@{}}Vulnerable\\Active\\Models\end{tabular} & 29 & 22 & 0 & 7 & 1 & 0 & \textbf{59} \\ 
\hline
\hline
\begin{tabular}[c]{@{}c@{}}Active\\Devices\end{tabular} & 246,184 & 983,832 & 38,293 & 3,102 & 742 & 3 & \textbf{1,272,156} \\ 
\hline
\rowcolor{mygray}
\begin{tabular}[c]{@{}c@{}}Vulnerable\\Active\\Devices\end{tabular} & 224,722 & 944,782 & 0 & 2,938 & 1 & 0 & \textbf{1,172,443} \\
\bottomrule[0.5pt]
\bottomrule
\end{tabular}
}
\end{table}

%% file: tables/dir-818lw.tex
\begin{table}[t]
    \centering
    \caption{The firmware information of {DIR-818LW\_REVA}.
    	The firmware {DIR-818LW\_REVA\_FIRMWARE\_1.01} does
    	not have build time because we cannot unpack the firmware.}
    \label{tab:818lw}
    \footnotesize
    \resizebox{0.35\textwidth}{!}{
        \begin{tabular}{c|cc}
        \toprule
        \toprule[0.5pt]
        Version & Release Date & Build Date \\
        \midrule
        1.01 & 01/24/2014 & - \\
        \rowcolor{mygray}
        1.02 & 03/27/2014 & 03/21/2014 \\
        1.04 & 09/01/2013 & 08/22/2014 \\
        \rowcolor{mygray}
        1.05 & 06/23/2015 & 04/19/2015 \\
        PATCH\_2.05 & 07/16/2015 & 07/09/2015 \\
        \rowcolor{mygray}
        \textbf{PATCH\_2.06} & \textbf{01/04/2019} & \textbf{12/06/2018} \\
        \bottomrule[0.5pt]
        \bottomrule
        \end{tabular}
    }
\end{table}

%% file: tables/vul_818.tex

\begin{table}[t]
    \centering
    \caption{Vulnerabilities exist in the {DIR-818LW\_REVA\_FIRMWARE\_1.05} firmware.}
    \label{tab:vul_818lw}
    \footnotesize
    \resizebox{0.55\linewidth}{!}{%
    \begin{tabular}{cc} 
    \toprule
    \toprule[0.5pt]
    Vulnerabilities & Risk Rank \\ 
    \midrule
    CVE-2020-15895 & Medium \\ 
    \rowcolor{mygray}
    CVE-2018-10106 & Critical \\ 
    CVE-2019-18852 & Critical \\ 
    \rowcolor{mygray}
    CVE-2018-6530 & Critical \\ 
    CVE-2019-20215 & Critical \\ 
    \rowcolor{mygray}
    CVE-2019–17621 & Critical \\
    \bottomrule[0.5pt]
    \bottomrule
    \end{tabular}
    }
    \end{table}

%% file: discussion.tex
\section{Discussion}
\label{sec:discussion}



First, our study raises alarms about EoL devices that deserve further efforts of the community.
For instance, Internet service providers can dynamically monitor the EoL devices and block attacks in the Internet.
Vendors can notify the owners of their EoL devices to upgrade the devices to newer ones.
A database of active EoL devices can be built and shared to let users detect whether their devices are EoL and provide security suggestions to users.
The community can also develop patches for critical vulnerabilities using binary rewriting techniques.
Nevertheless, our work is the first measurement study to reveal the alarming facts of EoL devices, making the community pay more attention to this issue that has been overlooked.

Second, we use the "app" filter in \code{ZoomEye} to collect the alive devices.
This method depends on the fingerprints that are used by the search engine to recognize models.
Although we have an evaluation showing that \code{ZoomEye} is more reliable than the other two engines, developing accurate fingerprints for every model is a time-consuming process because of the diversity of embedded devices.
Therefore, alive devices of some EoL models may be missed by the search engine. 
However, these cases are rare, and we think it is acceptable.

Third, although our study reveals the alarming facts about EoL devices, the result reported is still not the whole picture.
That is because our analysis is based on 894 EoL models of \code{D-Link}, \code{Netgear}, and \code{Tp-Link}, and there exist much more vendors in the market, e.g., Huawei and Cisco.
The findings may not directly apply to EoL models of other vendors.
Moreover, we get the EoL model lists of \code{D-Link} and \code{Tp-Link} from their US websites~\cite{dlink_eol_models}.
We notice that for different regions, EoL models can be different.
Therefore, we cannot claim our findings can apply to EoL models in other regions.
However, the methodology proposed by this study can be applied to other models, as long as the EoL information can be directly retrieved.

Fourth, our evaluation requires manual efforts to analyze security patches and the release notes of firmware images.
Although we are careful during this process, the involvement of manual efforts could introduce errors. 
In addition, we verify the existence of vulnerabilities in EoL models dynamically with \code{FIRMADYNE} or statically with \code{Ida pro} and \code{Ghidra}.
However, \code{FIRMADYNE} cannot emulate all the firmware images.
For the vulnerabilities that are statically verified through manual efforts, it is possible that the evaluation may have false positives and false negatives.
In the future, we may leverage other state-of-the-art firmware analysis systems if they are publicly available.

Last but not least, our methodology and experiment follow the basic ethical research guidelines~\cite{dittrich2012menlo}.
We carefully handle all collected data and do not involve human subjects during the experiment.
We will release the data about EoL models and vulnerabilities but not release the IP address of EoL devices to avoid providing information to attackers.

%% file: relatedwork.tex
\section{Related Work}
\label{sec:related}

\smallskip
\noindent \textbf{Studies of Embedded Systems}\tab
There are some large-scale studies of embedded systems.
Costin et al.~\cite{costin2014large} conduct the first large-scale analysis of embedded systems to assess their security and bring new insights into the security of embedded devices.
Bojinov et al.~\cite{Bojinov} conduct the first in-depth audit of embedded web management interfaces.
Cui et al.~\cite{cui2009brave} conduct a vulnerability assessment of embedded network devices within the world’s largest ISPs and civilian networks and find a large number of vulnerable embedded devices. 
Cui et al.~\cite{cui2010quantitative} scan the number of vulnerable embedded devices on a global scale and present the insecurity of embedded network devices.
HD Moore~\cite{upnp} reveals security flaws in UPnP (Universal Plug and Play) protocol and scans the affected devices.
Jiang et al.~\cite{jiang_empirical_2020} conduct the first comprehensive study on ARM disassembly tools to analyze their capabilities to locate instructions and function boundaries.
These studies concentrate on all the embedded devices while our study focuses on EoL devices.

\smallskip
\noindent \textbf{Security Analysis of Embedded Systems}\tab
Shoshitaishvili et al.~\cite{shoshitaishvili2016sok} present \code{angr}, a systematized implementation of various analysis techniques.
Chen et al.~\cite{chen2016towards} present \code{FIRMADYNE}, the first automated dynamic analysis system that can emulate Linux-based firmware of embedded devices in a scalable manner.
Zaddach et al.~\cite{zaddach2014avatar} present \code{Avatar} that enables complex dynamic analysis of embedded devices by orchestrating the execution of an emulator together with the real hardware.
Davidson et al.~\cite{davidson2013fie} present \code{FIE} that builds off the \code{KLEE} symbolic execution engine in order to provide an extensible platform for detecting bugs in firmware programs for the popular MSP430 family of microcontrollers.
Costin et al.~\cite{costin2016automated} present a fully automated framework to dynamically analyze the web interfaces in embedded systems.
Dullien et al.~\cite{dullien2005graph} present a method to construct an optimal isomorphism between the sets of instructions, sets of basic blocks, and sets of functions in two similar executables.
Ming et al.~\cite{ming2017binsim} perform enhanced dynamic slicing and symbolic execution to compare the logic of instructions that impact the observable behaviors to detecting differences between two binary executables.
These works promote the dynamic and static analysis techniques progress and help us to conduct the study.

\smallskip
\noindent \textbf{Patches for EoL Devices in Other Architectures}\tab
In other architectures, there is also no concept of End-of-Life, but there have been work noticed devices that are no longer supported by the manufacturer and mobile operator.
Mulliner et al.~\cite{mulliner2013patchdroid} present \code{PatchDroid}, a system to distribute and apply third-party security patches for Android.
For EoL embedded devices, it can also adopt similar methods to reduce the risk.